\documentstyle[pra,eqsecnum,aps]{revtex}

\def\displayfrac#1#2{\frac{\displaystyle #1}{\displaystyle #2}}
\begin{document}
\baselineskip15pt
\title{The Sampling Theorem and Coherent State Systems in Quantum Mechanics}
\author{ Arvind\thanks{email:arvind@physics.iitm.ac.in}\\ Department of
  Physics, 
Indian Institute of Technology Madras, Chennai 600 036, India\\
S. Chaturvedi\thanks{e-mail: scsp@uohyd.ernet.in}\\ 
 School of Physics, University of Hyderabad, Hyderabad 500 046,
India\\
N. Mukunda\thanks{email: nmukunda@cts.iisc.ernet.in}\\
 Centre for High Energy Physics,
 Indian Institute of Science, Bangalore 560 012, India\\
 R.Simon\thanks{email: simon@imsc.res.in}\\
  The Institute of 
 Mathematical Sciences, C. I. T. Campus, Chennai 600 113, India\\}
 \maketitle

\begin{abstract}
The well known Poisson Summation Formula is analysed from the perspective of the 
coherent state systems associated with the Heisenberg--Weyl group. In particular, it is shown 
that the Poisson summation formula may be viewed abstractly as a relation between  two sets of 
bases (Zak bases) arising as simultaneous eigenvectors of two commuting unitary operators 
in which geometric phase plays a key role. The Zak bases are shown to be interpretable as generalised coherent state systems of the Heisenberg--Weyl group and this, in turn, prompts analysis of the sampling theorem (an important and useful consequence of the Poisson Summation Formula) and its extension from a coherent state point of view  leading to interesting results 
on properties of von Neumann and finer lattices based on standard and generalised coherent state
systems.    
\end{abstract}

\newpage
\section{Introduction}

It is well known that the Sampling Theorem (ST) for band limited
signals\cite{1}, of fundamental importance in communication theory,
arises from the properties of the Fourier transform operation
on the real line, and the related Poisson Summation Formula. 
It is also well known that the Heisenberg-Weyl (H-W) group\cite{2}, 
which is basic for nonrelationistic quantum kinematics, 
is intimately related to the same Fourier transformation.  
This is clear from the 
description of particle momentum in wave mechanics, and in the 
position-momentum uncertainty principle.  As will become evident, 
it is possible to derive  the Poisson Summation Formula in a 
particularly elegant manner from the representation theory of
the H-W group.

Many applications of this group use the remarkable properties of 
the  so-called `coherent states' originally discovered by 
Schr\"odinger\cite{3}, and extensively used in quantum optics in 
particular\cite{4}. 
The theory of these and other systems of coherent states, called 
`generalized coherent states', has been put on a comprehensive 
footing, and the extension to such systems associated with general 
Lie groups has been carried out\cite{5}.  In the process it has been 
realised that even for a given Lie group, such as the H-W group, 
one can construct many different systems of generalised coherent 
states, sharing some features dictated by the structure of the
group, but differing from one another in certain details.

These remarks suggest that the H-W group functions as a unifying
element or as a common connecting thread linking various ideas 
and concepts, each of which figuratively flows out of the group
and its representations in a different direction - Poisson 
Summation Formula, Sampling Theorem, specific families of 
generalised coherent states and, as one finds, even certain
instances of the recently much studied geometric phase\cite{6}.
There is yet another sense in which the usual ST and the standard
coherent states share some common features.  There are certain
discrete subsets of the coherent states, namely the so-called 
von Neumann lattice of these states and finer lattices, which 
enjoy the property of `totality' or (over) completeness in the
relevant Hilbert space: any vector in this space is in principle
fully determined once one knows its inner products with all the 
vectors in the lattice\cite{7}. Evidently this too is in a sense a 
sampling theorem.  These lattices of states and some generalisations
have been studied extensively some time ago, developing in the
process simpler proofs of totality, analysis of conditions leading to 
orthonormality etc\cite{8}.  It would seem to be of 
considerable interest to
express the usual ST in such a way that a comparison with the
properties of lattices of coherent states, standard or
generalised, could be easily carried out.

 In this work we attempt to forge a certain sense of unity among 
 these various concepts from the perspective of coherent state systems 
 of the H--W group and seek extensions and generalisations of known 
 results to the extent possible. A brief outline of this work is as follows. 
 In Section II we recapitulate features of the H--W group to the extent 
 required in this work and show how the Poisson Summation Formula arises 
 as a consequence of the relation between two bases consisting of two 
 commuting unitary operators and highlight the role the geometric 
 phase plays in this context. In section III, we establish connection 
 between the two bases and the Zak representation \cite{11} and further show that the 
 two can be identified with certain generalised coherent states of the H--W 
 group We also discuss some of their special features needed later and in Section IV 
 give the Wigner distribution of the underlying fiducial vector. Section V 
 is devoted to two forms of the standard ST for band limited state vectors. In Section VI 
 we translate the contents of the standard ST into the properties of standard coherent state lattices and extend the results to a general state vector and compare them with 
 known results on von Neumann and finer standard coherent state lattices. Similar 
 questions in the context of generalised coherent state systems are explored in 
 in Section VII. Section VIII contains  concluding remarks and further outlook.  
       
\section{The H-W group and the Poisson Summation Formula}

The H-W group and its associated operator structures are based 
on the fundamental Heisenberg canonical commutation relation
     \begin{eqnarray}
     [\hat{q}, \hat{p}] = i
     \end{eqnarray}

\noindent
for hermitian operators $\hat{q}, \hat{p}$ representing position 
and momentum respectively for a one-dimensional Cartesian quantum
mechanical system.  (For simplicity we set Planck's constant 
$\hbar = 1$).  Thus this group is a three parameter Lie group whose
elements and composition law may be written as follows:
     \begin{mathletters}
     \begin{eqnarray}
     D(\alpha_1,\alpha_2,\alpha_3) &=&\exp\left\{
     -i\alpha_1\hat{p} + i\alpha_2\hat{q} -i\alpha_3
     \right\},\nonumber\\
     -\infty < \alpha_1, \alpha_2 < \infty &,&
     0\leq \alpha_3 < 2\pi ;\\
     D\left(\alpha^{\prime}_1,
\alpha^{\prime}_2,\alpha^{\prime}_3\right)
     D(\alpha_1, \alpha_2, \alpha_3) &=&
     D\left(\alpha^{\prime}_1+\alpha_1, \alpha^{\prime}_2+
     \alpha_2, \alpha^{\prime}_3+\alpha_3 \right.\nonumber\\
     &+&\left.\frac{1}{2}\left(\alpha_1^{\prime}\alpha_2 -
     \alpha^{\prime}_2\alpha_1\right)\right)
     \end{eqnarray}
     \end{mathletters}

\noindent
(In the element on the right, the final phase is understood to be 
taken modulo $2\pi$).  According to the Stone-von Neumann 
Theorem\cite{9} 
there is essentially only one nontrivial unitary irreducible
representation of this group, ie., only one irreducible 
hermitian representation of the commutation relation (2.1), 
apart from unitary equivalence.  We shall write ${\cal H}$ for the 
Hilbert space of this representation.

The displacement operators correspond to setting $\alpha_3=0$
and to taking $(\alpha_1,\alpha_2)$ to be a point $(q,p)$ in the
classical phase space or plane:
     \begin{eqnarray}
     D(q,p) = \exp\{i p \hat{q} -i q \hat{p}\}, \;-\infty
     <q,p<\infty .
     \end{eqnarray}

\noindent
Their basic properties are read off from eqn.(2.2):
     \begin{mathletters}
     \begin{eqnarray}
     D(q,p)^{-1} &=& D(q,p)^{\dag} = D(-q,-p) ;\\
     D(q^{\prime},p^{\prime}) D(q,p) &=& \exp
     \left\{\frac{i}{2}(p^{\prime}q-q^{\prime}p)\right\}
     D(q^{\prime}+q), p^{\prime}+p);\\
     D(q,p)^{-1} (\hat{q}\;\mbox{or}\; \hat{p}) D(q,p)&=&
     \hat{q} +q \;\mbox{or}\;\hat{p}+ p .
     \end{eqnarray}
     \end{mathletters}

\noindent
When $q$ or $p$ vanishes it is convenient to define
     \begin{eqnarray}
     U(p) &=& D(0,p) = e^{ip\hat{q}}, \nonumber\\
     V(q) &=& D(q,0) = e^{-iq\hat{p}}.
     \end{eqnarray}

\noindent
For these we have the useful relations
     \begin{eqnarray}
     D(q,p) &=& e^{iqp/2} V(q) U(p)\nonumber\\
     &=& e^{-iqp/2} U(p) V(q) ,\nonumber\\
     U(p) V(q)&=& e^{iqp} V(q) U(p).
     \end{eqnarray}

\noindent
This last relation for the unitary operators $U(p), V(q)$ is just 
the finite Weyl form of the commutation relation (2.1); the
phase factor present here is the geometric phase associated
with the H-W group.

Let us denote the usual delta function normalised ideal 
eigenvectors of $\hat{q}$ and $\hat{p}$, which form
continuous bases for ${\cal H}$, by angular and rounded
ket vectors respectively:
     \begin{eqnarray}
     \hat{q}|q>&=& q|q>, \hat{p}|p) = p|p), q,p\in {\cal R};
     \nonumber\\
     <q^{\prime}|q> &=&\delta (q^{\prime}-q),
     (p^{\prime}|p) = \delta(p^{\prime}-p);\nonumber\\
      <q|p)&=& \frac{1}{\sqrt{2\pi}}\;e^{iqp}.
     \end{eqnarray}

\noindent
On these the actions of the exponentiated unitary operators are:
     \begin{eqnarray}
     V(q)|q^{\prime}>&=&|q^{\prime}+q>,\nonumber\\
     D(q,p)|q^{\prime}>&=& e^{ip(q^{\prime}+q/2)}
     |q^{\prime}+q >;\nonumber\\
     U(p)|p^{\prime})&=& |p^{\prime} +p),\nonumber\\
     D(q,p)|p^{\prime})&=& e^{-iq(p^{\prime}+p/2)}|
     p^{\prime}+p).
     \end{eqnarray}

The operators $U(p)$ and $V(q)$ do not commute in general.
Now choose some real positive $q_0$ and write
     \begin{eqnarray}
     V_0 = V(q_0) = e^{-i q_0 \hat{p}}.
     \end{eqnarray}

\noindent
We ask for the smallest nontrivial value of $p$ in $U(p)$,
assumed positive, such that $U(p)$ commutes with $V_0$:
this happens for $p=2\pi/q_0$, so we define
     \begin{eqnarray}
     U_0 = U(2\pi/q_0) = e^{2\pi i\hat{q}/q_0},
     \end{eqnarray}

\noindent
and then  have
     \begin{eqnarray}
     U_0 V_0 = V_0 U_0.
     \end{eqnarray}

\noindent
It is important to observe that both unitary operators $U_0$
and $V_0$ are determined by the single parameter $q_0$.

We look for the simultaneous (ideal) eigenvectors of $U_0$ and 
$V_0$.  Their eigenvalues are phases which it is natural to 
parametrise as follows:
     \begin{eqnarray}
     U_0\rightarrow e^{2\pi iq/q_0}&\;,\;&
     q\in\left[-\frac{1}{2} q_0, \frac{1}{2} q_0\right] ;
     \nonumber\\
     V_0 \rightarrow e^{-i q_0 p}&\;,\;& p\in \left[
     -\pi/q_0, \pi/q_0\right].
     \end{eqnarray}

\noindent
Given a pair $(q,p)$ within these limits, ie., a point in the 
rectangle $R(q_0)$ in the phase plane with sides $q_0, 2\pi/q_0$
centred at the origin, we can build up a simultaneous (ideal) 
eigenvector of $U_0$ and $V_0$ either in the $|q>$ basis or in 
the $|p)$ basis.  For this we need to use the actions (2.8)
of $U$'s and $V$'s on these bases.  In this way we find after
elementary algebra:
     \begin{eqnarray}
     |q,p>&=& \displayfrac{q^{1/2}_0}{\sqrt{2\pi}}\;
     \sum\limits_{n\in{\cal Z}}\;e^{i n q_0 p}
     |q+n q_0> ,\nonumber\\
     (U_0\;\mbox{or}\;V_0) |q,p> &=&
      \left(e^{2\pi i q/q_0}\;\mbox{or}\; e^{-iq_0 p}\right)
      |q,p> ,\nonumber\\
      <q^{\prime},p^{\prime}| q,p>&=& \delta
      (q^{\prime}-q) \delta(p^{\prime}-p) .
       \end{eqnarray}

\noindent
This construction started from the eigenvectors $|q>$ of
$\hat{q}$.  Alternatively we can build up the simultaneous
eigenvectors starting from the basis $|p)$.  Then we find:
     \begin{eqnarray}
     |q,p)&=& q^{-1/2}_0\;\sum\limits_{n\in{\cal Z}}\;
     e^{-2\pi i n q/q_0} |p +2\pi n/q_0),\nonumber\\
     (U_0\;\mbox{or}\;V_0) |q,p) &=&\left(e^{2\pi i q/q_0}\;
     \mbox{or}\; e^{-i q_0 p}\right) |q,p) ,\nonumber\\
     (q^{\prime},p^{\prime}|q,p) &=& \delta(q^{\prime} -q)
     \delta(p^{\prime}-p).
     \end{eqnarray}

\noindent
We expect that these two solutions must be phase related.  
We easily find:
     \begin{eqnarray}
      <q^{\prime},p^{\prime}|q,p) = e^{iqp} \delta(q^{\prime}
      -q) \delta(p^{\prime} - p) ,
      \end{eqnarray}

\noindent
which implies
     \begin{eqnarray}
     |q,p) = e^{iqp} |q,p>
     \end{eqnarray}

\noindent
We will recognize in the next section that this phase is the 
same H-W geometric phase already present in eqn.(2.6).

The relation (2.16) in conjugate form is
     \begin{eqnarray}
     \displayfrac{q_0^{1/2}}{\sqrt{2\pi}}\;\sum\limits
     _{n\in{\cal Z}}\;e^{-i n q_0 p} <q+n q_0| =
     q^{-1/2}_0\; e^{iqp}\;\sum\limits_{n\in{\cal Z}}
     \;e^{2\pi i n q/q_0}(p+2\pi n/q_0| .
     \end{eqnarray}

\noindent
Let $|\psi>\in{\cal H}$ be a general normalisable vector 
with position and momentum space wavefunctions $\psi(q),
\varphi(p)$ respectively:
     \begin{eqnarray}
     \psi(q) = <q|\psi>&=&\displayfrac{1}{\sqrt{2\pi}}\;
     \int\limits^{\infty}_{-\infty} dp \varphi(p)
     e^{ipq},\nonumber\\
     \varphi(p) =(p|\psi>&=&\displayfrac{1}{\sqrt{2\pi}}\;
     \int\limits^{\infty}_{-\infty} dq \psi(q)
     e^{-iqp}.\nonumber\\
     \end{eqnarray}

\noindent
Then taking the products of the two sides of eqn.(2.17) with 
$|\psi>$  and reinstating the parameter ranges we get:
     \begin{eqnarray}
      q_0\;\sum\limits_{n\in{\cal Z}}\;e^{-i n q_0 p} \psi
      (q+n q_0) &=&\sqrt{2\pi}\;
      e^{i q p}\;\sum\limits_{n\in{\cal Z}}\;
      e^{2\pi i n q/q_0} \varphi(p+2\pi n/q_0) ,\nonumber\\
      q_0>0, && (q,p)\in R(q_0).
      \end{eqnarray}

\noindent
This is the Poisson Summation Formula for any Fourier transform 
pair $\psi(q),\varphi(p)$\cite{10}.  It is usually derived quite 
directly 
from the structure of the Fourier Series representation for a 
function of an angle variable, by extending it to a periodic
function on the full real line.  We see here that it arises very
naturally in a quantum mechanical context by constructing
simultaneous eigenvectors of the commuting unitary operators
$U_0, V_0$ in two ways and relating the results.  This brings out 
the connection to the H-W group.  We also see that extending 
$(q,p)$ in eqn.(2.19) outside $R(q_0)$ does not give any 
additional information.

\section{Connection to Zak representation as a Generalised
Coherent State System}

The simultaneous (ideal) eigenvectors of the commuting unitary 
operators $U_0$ and $V_0$ developed in two ways in the previous
section lead to new representations of vectors $|\psi>\in{\cal H}$,
distinct from the representations based on position and momentum 
wavefunctions $\psi(q)$ and $\varphi(p)$.  These are the Zak
representations of quantum mechanics\cite{11}, known and studied 
for a long time and exploited in particular to examine the von 
Neumann lattice of standard coherent states and its 
generalisations\cite{8}.  The states $|q,p>,|q,p)$ of eqns.(2.13,14)
are in fact the Zak basis states for ${\cal H}$.  We explore 
briefly in this Section the possibility of interpreting them
as a system of (ideal) generalised coherent states 
associated with the H-W group.  First we begin with the Zak 
representation in quantum mechanics.

Given $|\psi>\in{\cal H}$ with conventional wavefunctions 
$\psi(q), \varphi(p)$ where $q,p\in {\cal R}$, we define the
Zak wavefunction $\chi(q,p)$ of $|\psi>$ by
     \begin{eqnarray}
     \chi(q,p) &=& <q,p|\psi>\nonumber\\
     &=&\displayfrac{q^{1/2}_0}{\sqrt{2\pi}}\;
     \sum\limits_{n\in{\cal Z}}\;e^{-i n q_0
     p} \;\psi(q+n q_0).
     \end{eqnarray}

\noindent
(Here of course $q_0$ is a positive parameter chosen freely
and then held fixed).  This definition is based on eqn.(2.13).  
Equally well we can use eqn.(2.14) and define
     \begin{eqnarray}
     \tilde{\chi}(q,p) &=& (q,p|\psi>\nonumber\\
     &=& e^{-iqp} \chi(q,p)\nonumber\\
     &=&q_0^{-1/2}\;\sum\limits_{n\in {\cal Z}}\;
     e^{2\pi i n q/q_0}\varphi(p+2\pi n/q_0).
     \end{eqnarray}

\noindent
In both eqns.(3.1,2) it is understood that $(q,p)\in R(q_0)$.  
These equations define the so-called Zak transform, and exhibit 
the Hilbert space ${\cal H}$ as $L^2(R(q_0))$, in the sense
that for any $|\psi>\in{\cal H}$ we have
     \begin{eqnarray}
     <\psi|\psi>&=& \parallel |\psi>\parallel^2\nonumber\\
     &=&\int\limits_{\cal R} dq\; |\psi(q)|^2\nonumber\\
     &=& \int\limits_{\cal R} dp\; |\varphi(p) |^2\nonumber\\
     &=& {\int\int\atop {\scriptstyle{R(q_0)}}} dq\; dp
     \left(|\chi(q,p)|^2 \;\mbox{or}\;|\tilde{\chi}(q,p)|^2\right).
     \end{eqnarray}
 
\noindent
To recover all elements $|\psi>\in{\cal H}$ we must allow for 
all (Lebesgue) square integrable Zak wavefunctions 
$\chi(q,p)(\mbox{or}\;
\tilde{\chi}(q,p))$ over the phase space rectangle $R(q_0)$.  The 
inverse of the Zak transform expresses $\psi(q)$ and $\varphi(p)$ 
in terms of $\chi(q^{\prime},p^{\prime})$ and $\tilde{\chi}
(q^{\prime},p^{\prime})$:
     \begin{mathletters}
     \begin{eqnarray}
      \psi(q)&=& \displayfrac{q_0^{1/2}}{\sqrt{2\pi}}\;\int\limits
      ^{\pi/q_0}_{-\pi/q_0} dp\;e^{iqp}\;\tilde{\chi}([q],p),
       \nonumber\\
      q&=&[q]\;\mbox{mod}\; q_0\;,\;[q]\in\left(-\frac{1}{2}
       q_0, \frac{1}{2} q_0\right) ;\\
       \varphi(p)&=& q_0^{-1/2}\;\int\limits^{\frac{1}{2}q_0}
       _{-\frac{1}{2}q_0} dq\;e^{-iqp}\chi(q,[p]),\nonumber\\
       p&=&[p]\;\mbox{mod}\; 2\pi/q_0,\;[p]\in(-\pi/q_0, \pi/q_0).
       \end{eqnarray}
       \end{mathletters}

The Zak basis vectors have the following formal `periodicity' 
properties as are evident upon inspection from eqns.(2.13,14):
     \begin{mathletters}   
     \begin{eqnarray}
     |q+q_0, p> &=& e^{-iq_0p} |q,p>\nonumber\\
     |q,p+2\pi/q_0>&=& |q,p> ;\\
     |q+q_0,p)&=& |q,p) ,\nonumber\\
     |q,p+2\pi/q_0)&=& e^{2\pi iq/q_0} |q,p)
     \end{eqnarray}
     \end{mathletters}

\noindent
These differing behaviours of $|q,p>$ and $|q,p)$ are consistent 
with eqn.(2.16).
Indeed the geomeric phase factor appearing in eqn.(2.16) converts
strict periodicity with respect to $p$ and periodicity upto a phase
with respect to $q$ in the case of $|q,p>$, to exactly opposite 
properties for $|q,p)$.  The point to be now appreciated is that 
while for a general $|\psi>\in {\cal H}$ we have no conditions on 
$\chi(q,p)(\mbox{or}\;\tilde{\chi}(q,p))$ other than (Lebesgue)
square integrability over $R(q_0)$, if we restrict ourselves to 
a subset  of $|\psi>\in{\cal H}$ possessing continuous Zak 
wavefunctions we can say something specific.  Namely, based on 
eqn.(3.5) we have for such vectors in ${\cal H}$ the properties
       \begin{mathletters}
       \begin{eqnarray}
       \chi\left(\frac{1}{2} q_0, p\right) &=&
       e^{i q_0 p}\chi\left(-\frac{1}{2} q_0,p\right),\nonumber\\
       \chi(q,\pi/q_0)&=& \chi(q,-\pi/q_0);\\
       \tilde{\chi}\left(\frac{1}{2} q_0,p\right)&=& 
       \tilde{\chi}\left(-\frac{1}{2} q_0, p\right) ,\nonumber\\
       \tilde{\chi}(q,\pi/q_0)&=& e^{-2\pi iq/q_0}
       \tilde{\chi}(q,-\pi/q_0).
       \end{eqnarray}
       \end{mathletters}

\noindent
For such vectors $|\psi>\in{\cal H}$ these relations among the 
values of the Zak wave functions along the edges of $R(q_0)$ 
can be exploited to show that $\chi(q,p)(\mbox{or}\;\tilde{\chi}(q,p))$ 
must have at least one zero in their domain of definition\cite{12}.

It is worth remarking that on account of the robustness of 
geometric phases, namely the impossibility of transforming them 
away by using phase redefinitions permitted by quantum mechanics,
we cannot replace eqns.(3.6) in any natural way by some related
wave functions over $R(q_0)$ strictly periodic simultaneously in 
both $q$ and $p$.  At best the geometric phase $e^{iqp}$ of 
eqn.(2.16) can be shifted from one place to another; and its 
presence is the essential reason behind the interesting result
mentioned in the previous paragraph.

It may be of interest to see briefly how the original operators
$\hat{q},\hat{p}$ obeying the commutation relation (2.1) act on
the Zak wavefunctions.  It turns out that in both cases we have
to restrict the wavefunctions $\chi, \tilde{\chi}$ to be continuous
 and once differentiable in each argument (so that the periodicity 
conditions (3.6) do apply) and then we have:
     \begin{mathletters}
     \begin{eqnarray}
     \chi(q,p)&:&\hat{q}=q+i\frac{\partial}{\partial p}\;,\;
     \hat{p} = -i\frac{\partial}{\partial q} \;;\\
     \tilde{\chi}(q,p) &:& \hat{q} = i\frac{\partial}{\partial p}\;,\;
     \hat{p} = p -i\frac{\partial}{\partial q} .
     \end{eqnarray}
     \end{mathletters}

After this brief recollection of the Zak representation of quantum
mechanics, we turn to the possibility of viewing the Zak basis 
states as an (ideal) system of generalised coherent states with
respect to the H-W group.  At $q=p=0$ the connection (2.16)
simplifies and we are led to define
     \begin{eqnarray}
     \Psi_0 = |0,0> = |0, 0),
     \end{eqnarray}

\noindent
it being understood that this is not a normalisable vector in 
${\cal H}$.  Now eqns.(2.13,14) show us how to build up 
$|q,p>$ and $|q,p)$ from $\Psi_0$ in natural ways using
the displacement operators in eqn. (2.3):
     \begin{mathletters}
     \begin{eqnarray}
     |q,p>&=&\displayfrac{q_0^{1/2}}{\sqrt{2\pi}}\;
     \sum\limits_{n\in{\cal Z}}\;e^{inq_0p} |q+n q_0>\nonumber\\
     &=&V(q) \displayfrac{q_0^{1/2}}{\sqrt{2\pi}}\;
     \sum\limits_{n\in{\cal Z}}\;e^{inq_0p} |nq_0>
     \nonumber\\
     &=&V(q)\;U(p)\;\Psi_0\nonumber\\
     &=&e^{-iqp/2} D(q,p)\Psi_0 ;\\
     |q,p)&=& q_0^{-1/2}\;\sum\limits_{n\in{\cal Z}}\;
     e^{-2\pi i n q/q_0} |p+2\pi n/q_0)\nonumber\\
     &=& q^{-1/2}_0 U(p) \;\sum\limits_{n\in{\cal Z}}\;
     e^{-2\pi i n q/q_0}|2\pi n/q_0)\nonumber\\
     &=&U(p)\;V(q)\;\Psi_0\nonumber\\
     &=& e^{iqp/2} D(q,p) \Psi_0
     \end{eqnarray}
     \end{mathletters}

\noindent
This shows that the simultaneous eigenvectors of $U_0$ and $V_0$
actually form the orbit  of the (ideal) fiducial vector
$\Psi_0$ under the H-W group (save for the phase factor in
$D(\alpha_1,\alpha_2,\alpha_3)$).  Incidentally eqn.(3.9)
show explicitly that the phases $e^{iqp}$ appearing in 
eqns.(2.6,16) have a common origin.

To identify the Zak basis vectors as a family of generalized
coherent states, at least in a formal sense, we must identify
within the H-W group the stability group of the fiducial vector
$\Psi_0$.  From eqns.(2.13,2.14,3.8) we have the obvious 
properties
     \begin{eqnarray}
     U_0\; \Psi_0 = V_0\;\Psi_0 = \Psi_0 ,
     \end{eqnarray}

\noindent
which lead to the invariances of the Zak basis states in the form
     \begin{eqnarray}
      D(q\pm q_0, p)\Psi_0 &=&
      e^{\mp i q_0 p/2}D(q,p) \Psi_0 ,\nonumber\\
      D(q,p\pm 2\pi/q_0)\Psi_0 &=&
      e^{\pm i\pi q/q_0} D(q,p) \Psi_0 .
      \end{eqnarray}

\noindent
Thus the stability subgroup of $\Psi_0$ within the H-W group 
is an infinite discrete abelian subgroup $H_{q_{0}}$
generated by $U_0$ and $V_0$:
     \begin{eqnarray}
     H_{q_{0}} &=&
     \left\{e^{-i\alpha_3+i\alpha_2 \hat{q} -i\alpha_1\hat{p}}
     |\alpha_3=0, \alpha_1=n q_0, \alpha_2 =2\pi m/q_0,
     m,n\in{\cal Z}\right\}\nonumber\\
     &=&\left\{U_0^m V_0^n |m,n\in{\cal Z}\right\} .
     \end{eqnarray}

\noindent
This means that the orbit of $\Psi_0$, namely the collection
of states $\{|q,p>\}$ say, is essentially the coset space of
the H-W group with respect to $H_{q_{0}}$.  This is identifiable
with the  rectangle $R(q_0)$ of area $2\pi$ in phase space, and 
so we see again in a natural way why we may limit $(q,p)$ to 
this rectangle in the Zak representation.

To sum up, the simultaneous (ideal) eigenvectors of $U_0$ and
$V_0$ form a system of generalised coherent states for the H-W
group, based on the fiducial vector $\Psi_0$ and identifiable
with the coset space (H-W group)/$H_{q_{0}}$.  We must however 
note the following: Unlike the usual cases of generalised coherent 
state systems arising from a fiducial vector which is a 
normalisable vector in ${\cal H}$, in which case the inner
product of two generalised coherent states is generally 
nonzero\cite{13}, here we have
     \begin{eqnarray}
     \left(D(q^{\prime},p^{\prime})\Psi_0\;,\;
     D(q,p)\Psi_0\right) =
     \delta(q^{\prime}-q) \delta(p^{\prime}-p).
     \end{eqnarray}

\noindent
We realise that this result of orthonormality in the continuous 
Dirac sense is possible only because $\Psi_0$ is nonnormalisable.

\section{Wigner distribution for Zak fiducial vector}

The important role played by the fiducial vector $\Psi_0$ 
motivates us to explore its invariances in the Wigner 
representation language, more particularly since the
primitive invariances are with respect to phase space
displacements.  The position and momentum space wavefunctions 
of $\Psi_0$ are:
     \begin{eqnarray}
     \Psi_0(q) = <q|\Psi_0> &=& \displayfrac{q_0^{1/2}}
     {\sqrt{2\pi}}\;\sum\limits_{n\in{\cal Z}}\;
     \delta(q-n q_0),\nonumber\\
     \Phi_0(p) = (p|\Psi_0>&=&
     q_0^{-1/2}\;\sum\limits_{n\in{\cal Z}}\;
     \delta(p-2\pi n/q_0),\;q,p\in{\cal R} .
     \end{eqnarray}

\noindent
Notice that in both cases we have a periodic sequence of delta
functions with uniform positive weights.  Each of these is quite 
easily seen to display the basic invariances (3.10) of $\Psi_0$.
>From here we obtain the Wigner function corresponding to 
$\Psi_0$:
     \begin{eqnarray}
     W_0(q,p) &=& \frac{1}{2\pi}\int\limits^{\infty}_{-\infty}
     dq^{\prime}\;\Psi_0\left(q-\frac{1}{2}q^{\prime}\right)
     \Psi_0\left(q+\frac{1}{2}q^{\prime}\right)^*
     e^{iq^{\prime}p}\nonumber\\
     &=&\frac{q_0}{(2\pi)^2}\;\sum\limits_{n,n^{\prime}\in{\cal Z}}
     \int\limits^{\infty}_{-\infty} dq^{\prime} \delta\left(
     q-\frac{1}{2}q^{\prime}-nq_0\right) \delta\left(
     q+\frac{1}{2} q^{\prime}-n^{\prime}q_0\right) e^{iq^{\prime}p}
     \nonumber\\
     &=&\displayfrac{q_0}{(2\pi)^2}\;\sum\limits_{n,n^{\prime}
     \in{\cal Z}}\;\delta\left(2q-\left(n+n^{\prime}\right)q_0\right)
     \int\limits^{\infty}_{-\infty} dq^{\prime}\;
     e^{iq^{\prime}p}\;\delta\left(q-\frac{1}{2}q^{\prime}-nq_0
     \right)\nonumber\\
     &=&\displayfrac{q_0}{(2\pi)^2}\;\sum\limits_{n,n^{\prime}\in
     {\cal Z}}\;\delta\left(q-\frac{1}{2}\left(n+n^{\prime}\right)
     q_0\right)\;e^{2 i p(q-nq_0)}\nonumber\\
     &=&\displayfrac{q_0}{(2\pi)^2}\;e^{2 i q p}\;\sum\limits
     _{m,n\in{\cal Z}}\;\delta\left(q-\frac{1}{2} m q_0\right)
     e^{-2i n p q_0}\nonumber\\
     &=&\displayfrac{q_0}{(2\pi)^2}\;e^{2 i  q p}\;\sum\limits
     _{m\in{\cal Z}}\;\delta\left(q-\frac{m}{2} q_0\right)\;
     2\pi\frac{1}{2q_0}\cdot\sum\limits_{n\in{\cal Z}}
      \;\delta(p-\pi n/q_0)\nonumber\\
      &=& \frac{1}{4\pi}\;\sum\limits_{m,n\in{\cal Z}}\;(-1)^{mn}\;
      \delta\left(q-\frac{m}{2} q_0\right)\;
      \delta(p-n\pi/q_0).
      \end{eqnarray}

\noindent
We have here a lattice of delta functions in the $q-p$ phase plane, 
at the points $(\frac{m}{2} q_0, n\pi/q_0)$ for all $m,n\in{\cal Z}$.  
Thus the lattice spacings are $\frac{1}{2} q_0$ along the 
$q$-axis and $\pi/q_0$ along the $p$-axis.  The primitive 
cell here is one-fourth 
of $R(q_0)$ encountered earlier in constructing the 
$U_0-V_0$ eigenstates.  When $m$ and $n$ are both odd we 
have weight - 1, 
otherwise always weight +1.  This makes the invariances
     \begin{eqnarray}
     W_0(q\pm q_0, p) = W_0(q,p \pm 2\pi/q_0) =
     W_0(q,p)
     \end{eqnarray}

\noindent
immediately obvious. It is interesting to note that this (idealised) Wigner function and 
its properties are reminiscent of the Talbot grating in classical wave optics.

\section{The Sampling Theorem for band limited wavefunctions}

The results so far discussed have depended on one positive 
parameter $q_0$ with dimension of length.  We now turn to 
results which depend in addition on a second (positive)
parameter $p_0$ with dimension of momentum, such that
$p_0\leq 2\pi/q_0$.  The first is the ST recalled in this
Section \cite{1}.  The second, taken up in the next Section, is the 
property of (over) completeness possessed by certain lattices 
of standard quantum mechanical (Schr\"odinger) coherent states.

Let $|\psi>\in{\cal H}$ be such that its momentum space wavefunction 
$\varphi(p)$ vanishes for $p$ outside the interval 
$\left[-\frac{1}{2} p_0, \frac{1}{2} p_0\right]$ of width $p_0$, for
some positive $p_0$.  Therefore
     \begin{eqnarray}
     \psi(q) = \frac{1}{\sqrt{2\pi}}\;
      \int\limits^{\frac{1}{2} p_0}
      _{-\frac{1}{2} p_0} dp\;e^{ipq} \varphi(p).
     \end{eqnarray}

\noindent
We will then say that $\psi$ is band limited and has bandwidth
$p_0$.  (Conventionally the bandwidth is the length of the
smallest closed interval, centred about zero, outside of
which $\varphi(p)$ vanishes; however in the present context
it is more convenient to use the above definition, without
insisting that $\varphi(p)$ be nonzero throughout the interval
$\left[-\frac{1}{2} p_0, \frac{1}{2} p_0\right]$).  It is then 
useful to define a subspace ${\cal H}_0(p_0)\subset {\cal H}$, 
made up of band limited $\psi(q)$ with band width $p_0$, as
follows:
     \begin{eqnarray}
     {\cal H}_0(p_0) = \left\{|\psi>\in{\cal H}\bigg|
     \varphi(p)=0\;\mbox{for}\; p\;{\not\in} 
     \left[-\frac{1}{2} p_0, \frac{1}{2} p_0\right]\right\} 
     \subset{\cal H}.
      \end{eqnarray}

\noindent
(The reason for the subscript zero will become clear in the
following Section).  We can now see that if $p_0\leq 2\pi/q_0$, 
the subspace ${\cal H}_0(p_0)$ is very simply characterised in 
terms of Zak wavefunctions, namely as is clear from eqns.
(3.2,4):
     \begin{eqnarray}
     |\psi>\in{\cal H}_0 (p_0)\;,\; p_0\leq 2\pi/q_0
      &\Leftrightarrow & \tilde{\chi}(q,p) = \tilde{\chi}(p)
     \;\mbox{independent of}\; q ,\nonumber\\
     \varphi(p)&=& q_0^{1/2} \tilde{\chi}(p)\;, \;
     p\in\left[-\frac{1}{2} p_0, \frac{1}{2} p_0\right]
     \subseteq [-\pi/q_0, \pi/q_0] .
      \end{eqnarray}

\noindent
We will hereafter regard $q_0$ as given right at the start 
and kept fixed, so that the domain of definition of Zak wavefunctions 
$\chi(q,p),\tilde{\chi}(q,p)$ is the rectangle $R(q_0)$ in 
phase space, and this is unvarying.  The second parameter
$p_0$ will be permitted to vary subject always to 
$p_0\leq 2\pi/q_0$.  With this understood, the relation (5.3) 
discloses a natural connection between position -independent
Zak wavefunctions $\tilde{\chi}(q,p)$ and band limited wave 
functions $\psi(q)$ with band width $p_0$.

Now we proceed to the ST   We appeal to the Poisson Summation
Formula (2.19) which holds for any $q_0>0$.  For given band limited 
$\psi(q)$ with band width $p_0\leq 2\pi/q_0$, ie. 
$|\psi>\in {\cal H}_0(p_0)$, the interval 
$\left[-\frac{1}{2} p_0, \frac{1}{2} p_0
\right]$ does not extend beyond the interval 
$\left[-\pi/q_0, \pi/q_0\right]$.  
If we now take $p\in\left[-\frac{1}{2} p_0,
\frac{1}{2}p_0\right]$, all the conditions for the validity of
eqn.(2.19) are obeyed and furthermore only the term $n=0$
survives on the right hand side of that equation.  Therefore
for $|\psi>\in{\cal H}_0(p_0), p_0\leq 2\pi/q_0$, we have:
     \begin{eqnarray}
     \varphi(p)&=&\displayfrac{q_0}{\sqrt{2\pi}}\;
     e^{-iq^{\prime} p}\;\sum\limits_{n\in{\cal Z}}\;
     e^{-i n q_0 p}\;\psi(q^{\prime}+n q_0),\nonumber\\
     q^{\prime}\in\left[-\frac{1}{2} q_0, \frac{1}{2} q_0
      \right]&,& p\in \left[-\frac{1}{2} p_0, \frac{1}{2} 
     p_0\right] \subseteq \left[-\pi/q_0,\pi/q_0\right].
     \end{eqnarray}

\noindent
Using this in eqn.(2.18) we are able to express $\psi(q)$ for
any $q\in{\cal R}$ in terms of the discrete equispaced sequence 
of values $\psi\left(q^{\prime}+n q_0\right)$:
     \begin{eqnarray}
     \psi(q) &=& \frac{q_0}{2\pi} \;\sum\limits_{n\in{\cal Z}}
     \;\psi\left(q^{\prime}+n q_0\right)\;
     \int\limits^{\frac{1}{2} p_0}_{-\frac{1}{2} p_0} dp
     \;e^{ip\left(q-q^{\prime}-n q_0\right)}\nonumber\\
     &=&\frac{q_0}{\pi}\;\sum\limits_{n\in{\cal Z}}\;
     \displayfrac{\sin\left\{p_0\left(q-q^{\prime}-n q_0\right)
     /2\right\}}{\left(q-q^{\prime}-n q_0\right)}\;\;
    \psi\left(q^{\prime} + n q_0\right),\nonumber\\
    q\in{\cal R}&,& q^{\prime}\in\left[-\frac{1}{2} q_0, 
      \frac{1}{2} q_0 \right],\; \;p_0\leq 2 \pi/q_0 .
    \end{eqnarray}
     
\noindent
This is, as is well known, the ST for band limited $\psi(q)$.  
However in the usual statement, the band width $p_0$ is 
supposed to be known, and the inequality $p_0\leq 2\pi/q_0$ is
read as $q_0\leq 2\pi/p_0$ and taken to mean that the values of
$\psi(q^{\prime}+nq_0)$ are needed at sufficiently close
spacing in order to be able to determine $\psi(q)$ for all
$q$.

If in eqn.(5.5) we let $q\rightarrow q^{\prime} +m q_0$ for some 
$m\in{\cal Z}$, we find:
     \begin{eqnarray}
     \psi(q^{\prime} +m q_0) &=&\frac{q_0}{\pi}
     \left(\frac{p_0}{2} \psi(q^{\prime}+m q_0) +
     \sum\limits_{{n\in{\cal Z}\atop n\neq m}}\;
     \displayfrac{\sin\{(m-n)q_0 p_0/2\}}
     {(m-n)q_0}\;\psi(q^{\prime}+n q_0)\right) ,\nonumber\\
     && q^{\prime} \in \left[-\frac{1}{2} q_0,
      \frac{1}{2} q_0\right].
     \end{eqnarray}

\noindent
For $p_0< 2\pi/q_0$ this shows that the values of $\psi(q)$ at 
the discrete set of points $q^{\prime}+n q_0$, while certainly 
adequate 
to determine $\psi(q)$ in its entirety, can not be chosen 
independently.  There are linear relations among them, and more 
such relations will be described below.  For $p_0=2\pi/q_0$, 
eqn.(5.6) becomes an identity.  

In the form (5.5) for the ST, when $p_0< 2\pi/q_0$, the band
width $p_0$ appears explicitly on the right hand side.  It is
interesting that there is an alternative derivation and
expression of the ST, based on Cauchy's theorem for
analytic functions, in which $p_0$ does not appear explicitly but 
only implicitly.  From eqn.(5.1) it is evident that $\psi(q)$ 
is the boundary value, on the real axis, of an entire analytic 
function $\psi(z)$ defined for all $z\in{\cal C}$ by
     \begin{eqnarray}
     \psi(z) = \frac{1}{\sqrt{2\pi}} \;\int\limits^{\frac{1}{2}
     p_0}_{-\frac{1}{2} p_0}  dp\;e^{i p z} \;\varphi(p).
     \end{eqnarray}

\noindent
Whereas, by the Riemann-Lebesgue lemma, as $q\rightarrow \pm 
\infty$ along the real axis $\psi(q)$ definitely tends to zero,
we now see from the band limitedness that as $|z|\rightarrow
\infty$ in the complex plane the behaviour of $\psi(z)$ is 
controlled by
     \begin{eqnarray}
      |\psi(z)| \leq\; \mbox{constant}\;
      \exp\left(\frac{1}{2}
      p_0 |\;\mbox{Im}\;z|\right).
      \end{eqnarray}

\noindent
Now, for fixed $q^{\prime}\in\left[-\frac{1}{2} q_0, 
\frac{1}{2} q_0\right]$, set up the analytic function
     \begin{eqnarray}
     f(z) =\displayfrac{\pi}{\sin \pi z/q_0}\;
      \displayfrac{\psi(q^{\prime} + z)}{q^{\prime} + z-z_0} ,
     \end{eqnarray}

\noindent
where $z_0\in{\cal C}$ with $\mbox{Im}\;z_0\neq 0$.  This function
has simple poles at $z=z_0-q^{\prime}$ and $z=n q_0, n\in
{\cal Z}$.  As $|z|\rightarrow\infty$, on account of (5.8)
$|f(z)|$ tends to zero exponentially rapidly (and for this we do 
need the strict inequality $p_0< 2\pi/q_0$).  Thus using Cauchy's 
residue theorem for a contour consisting of a circle of large
radius centred at the origin, and letting the radius tend to
infinity, we get the result
     \begin{eqnarray}
     \psi(z_0)&=& \frac{q_0}{\pi}\;\sin\left\{\pi
     \left(z_0-q^{\prime}\right)/q_0\right\}\;
     \sum\limits_{n\in{\cal Z}}\;
     (-1)^n\;\displayfrac{\psi\left(q^{\prime}
     +n q_0\right)}{\left(z_0-q^{\prime}-n q_0\right)} .
     \end{eqnarray}

\noindent
We now let $z_0\rightarrow q\in{\cal R}$ to finally get:
     \begin{eqnarray}
     \psi(q) &=& \frac{q_0}{\pi}\;\sin
      \left\{\pi(q-q^{\prime})/q_0\right\}\;
      \sum\limits_{n\in{\cal Z}}\;
      (-1)^n\;\displayfrac{\psi\left(q^{\prime}
     +n q_0\right)}{\left(q-q^{\prime}-n q_0\right)} ,
      \nonumber\\
      && q^{\prime}\in\left[-\frac{1}{2} q_0, 
      \frac{1}{2} q_0\right],\;p_0< 2\pi/q_0 .
     \end{eqnarray}

\noindent
This differs in structure and properties from eqn.(5.5).  As
mentioned  earlier, the band width $p_0$ is not explicitly 
present on the right hand side; and as $q\rightarrow q^{\prime}
+m q_0$ for some $m\in{\cal Z}$, we get an identity rather 
than a nontrivial relation like (5.6).  The fact that the values of
$\psi(q^{\prime}+n q_0),\;n\in{\cal Z}$, are not all
independent when $p_0< 2\pi/q_0$ permits the existence of
both eqns.(5.5,11) having somewhat different forms.  It is
interesting to notice that even though we assumed 
$p_0<2\pi/q_0$ in the Cauchy theorem derivation of eqn.(5.11),
if we do take $p_0=2\pi/q_0$ the two results (5.5,11) become
identical.

To show even more forcefully, when $p_0< 2\pi/q_0$, that 
$\psi\left(q^{\prime}+n q_0\right)$ for $n\in {\cal Z}$ are not 
all independent, consider in place of $f(z)$ of eqn.(5.9)
the analytic function
     \begin{eqnarray}
     g(z) =\frac{\pi}{\sin \pi z/q_0}\cdot
     \psi(q^{\prime}+z)\;P(z),
     \end{eqnarray}

\noindent
where $P(z)$ is any finite degree polynomial.  The conditions for 
the use of Cauchy's theorem for the same circular contour as
before, and going to the limit of infinite radius, are all
obeyed.  In that limit we get the result
     \begin{eqnarray}
     \sum\limits_{n\in{\cal Z}} (-1)^n P(n q_0) \psi
     \left(q^{\prime} + nq_0\right) = 0 .
     \end{eqnarray}

\noindent
Thus we have infinitely many such linear dependence relations, 
the independent ones among them corresponding to choosing
$P(z)$ to be any monomial $z^m, m\in{\cal Z}$.  The important
point is that in the above argument $P(z)$ must be a polynomial 
of finite degree.  If it were a nontrivial entire function, 
its behaviour as $|z|\rightarrow \infty$ could spoil the behaviour 
of $g(z)$ and then Cauchy's theorem becomes inapplicable in 
general.

\section{Extended Sampling Theorem and Standard Coherent State
lattices}

We have mentioned in the Introduction that certain well-known 
theorems pertaining to phase space lattices of the standard 
coherent states in quantum mechanics have a character very 
similar to the ST discussed in the preceding Section.
Furthermore the Zak representation of quantum mechanical wave
functions has proven very useful in understanding (at least)
the von Neumann lattice of standard coherent states, and
in posing the problem of generalising this lattice\cite{8}. In
the present Section we combine the usual statement of the ST 
with the operator machinery provided by the H-W group to 
find the maximum extent to which the ST can be generalised
and expressed in terms of the standard coherent states.
Thus our aim is to see if the ST can be extended from vectors 
$|\psi>\in {\cal H}_0(p_0)$ to all $|\psi>\in{\cal H}$.  We 
then state the known results about lattices of standard 
coherent states, and show how close the two results are in
appearance and exactly where they differ.

We first recall briefly the definition and wave functions of 
the standard coherent states\cite{2,4}, the actions of the phase 
space displacement operators on them, and an interesting way in
which certain coherent states can be obtained from the
(ideal) position and momentum eigenvectors $|q>$ and $|p)$.
With this preparation we are able to recast and extend
the ST in the language of phase space lattices of
coherent states.

The standard coherent states are labelled by complex
numbers $z\in{\cal C}$; for clarity they will be written
as $|z))$.  Their definition in terms of the H-W
displacement operators and their wave functions are:
     \begin{mathletters}
     \begin{eqnarray}
     z&=&\frac{1}{\sqrt{2}} (q+ i p) :\nonumber\\
     |z))&=& |\frac{1}{\sqrt{2}} (q+ip)))\nonumber\\
     &=& D(q,p) |0))\nonumber\\
     &=&e^{\frac{i}{2}qp} V(q) U(p) |0))\nonumber\\
     &=&e^{-\frac{i}{2} qp} U(p) V(q) |0)) ;\\
     <q^{\prime}|z))&=& \frac{1}{\pi^{1/4}}\exp
     \left\{-\frac{i}{2} qp + i p q^{\prime} - \frac{1}{2}
     (q^{\prime}-q)^2\right\},\nonumber\\
     (p^{\prime}|z))&=& \frac{1}{\pi^{1/4}} \exp
     \left\{\frac{i}{2} q p - i q p^{\prime} -\frac{1}{2}
     (p^{\prime} - p)^2\right\}.
     \end{eqnarray}
     \end{mathletters}

\noindent
These states are normalised to unity and no two of them are
mutually orthogonal.  They are (right) eigenstates of the 
annihilation operator $\hat{a}$:
     \begin{eqnarray}
     \hat{a} = \frac{1}{\sqrt{2}} (\hat{q} + i\hat{p}) :
     \hat{a}|z)) = z|z)).
     \end{eqnarray}

\noindent
The actions of $V(q^{\prime})$ and $U(p^{\prime})$ 
are easily obtained:
     \begin{eqnarray}
     V(q^{\prime})|\frac{1}{\sqrt{2}} (q+ i p)))&=&
     e^{-\frac{i}{2} pq^{\prime}}|\frac{1}{\sqrt{2}}
     (q+q^{\prime} +ip))) ,\nonumber\\
     U(p^{\prime})|\frac{1}{\sqrt{2}}(q+ip))) &=&
     e^{\frac{i}{2} qp^{\prime}}|\frac{1}{\sqrt{2}}
     (q+ip+ip^{\prime}))).
     \end{eqnarray}

It is interesting that particular cases of these coherent 
states can be obtained from the ideal vectors $|q>$ and $|p)$ 
by application of certain bounded hermitian operators to 
them\cite{14}. Define two operators $S_1, S_2$ on ${\cal H}$ by
     \begin{eqnarray}
     S_1 = e^{-\frac{1}{2}\hat{q}^2}\;,\;
     S_2 = e^{-\frac{1}{2}\hat{p}^2}.
     \end{eqnarray}

\noindent
It is clear that they are both hermitian and bounded, while
their inverses are hermitian and unbounded.  Under
similarity transformations applied respectively to
$\hat{p}$ and to $\hat{q}$ we find:
     \begin{mathletters}
     \begin{eqnarray}
     S_1 \hat{p} S_1^{-1}&=& -i\sqrt{2} \hat{a},\\
     S_2 \hat{q} S_2^{-1}&=& \sqrt{2} \hat{a} .
     \end{eqnarray}
     \end{mathletters}

\noindent
Therefore $S_1|p)$ and $S_2|q>$ are particular coherent
states $|z))$.  We find upon checking their wavefunctions
that 
     \begin{mathletters}
     \begin{eqnarray}
     S_1|p)&=&\frac{1}{\pi^{1/4}\sqrt{2}}|\frac{i}{\sqrt{2}}
     p)) ,\\
     S_2|q>&=& \frac{1}{\pi^{1/4}{\sqrt2}}|\frac{1}
     {\sqrt{2}} q)) .
     \end{eqnarray}
     \end{mathletters}

\noindent
Hereafter we mainly exploit eqn.(6.6b).  On the basis of 
these relations we can express the content of the ST, 
eqns.(5.5,11), in an equivalent way in the language
of these coherent states.

In eqn.(5.2) we have defined the subspace 
${\cal H}_0(p_0)\subset {\cal H}$ consisting of 
band limited wavefunctions $\psi(q)$ with band width $p_0$.
Clearly ${\cal H}_0(p_0)$ is invariant under action by
$S_2$, and moreover when restricted to ${\cal H}_0(p_0)$
the inverse $S^{-1}_2$ is also bounded.  Now the content
of the ST may be expressed in this way: given $q_0$
to begin with, ensuring $p_0\leq 2\pi/q_0$ and
choosing $q^{\prime}\in\left[-\frac{1}{2} q_0,
\frac{1}{2} q_0\right]$,
     \begin{eqnarray}
     |\psi>\in {\cal H}_0(p_0),\;\psi\left(q^{\prime}+nq_0\right)
     =0,\;\mbox{all}\;n\in{\cal Z}\Rightarrow |\psi>=0 .
     \end{eqnarray}

\noindent
In other words such a band limited $|\psi>$ is (possibly over) 
determined by the values of $\psi\left(q^{\prime}+nq_0\right)$
for fixed $q^{\prime}$ and all $n\in{\cal Z}$.  For simplicity
now set $q^{\prime}=0$.  Then the ST is equivalent to the 
statement
     \begin{eqnarray}
     |\psi>\in{\cal H}_0 (p_0)\;,\;<nq_0|\psi>=0,\;\mbox
     {all}\;n\in{\cal Z}\Rightarrow |\psi>=0 .
     \end{eqnarray}

\noindent
The interesting aspect of this statement is that the 
(ideal) vectors $|nq_0>$ are in no sense vectors in 
${\cal H}_0(p_0)$, though they of course have nonzero
projections on to ${\cal H}_0(p_0)$.  Now from the above 
mentioned properties of $S_2$ with respect to 
${\cal H}_0(p_0)$ we have on the one hand
     \begin{eqnarray}
     |\psi>\in{\cal H}_0(p_0)\Longleftrightarrow
     S_2|\psi>, S_2^{-1} |\psi> \in
     {\cal H}_0(p_0) ;
     \end{eqnarray}

\noindent
and on the other hand
     \begin{eqnarray}
     S_2 |nq_0 > &=& \frac{1}{\pi^{1/4}\sqrt{2}}|\frac{1}
     {\sqrt{2}} n q_0)) ,\nonumber\\
     <nq_0|S_2 &=& \frac{1}{\pi^{1/4}\sqrt{2}}
     ((\frac{1}{\sqrt{2}} n q_0|.
     \end{eqnarray}

\noindent
Combining these facts we see that the ST is equivalent to
the following claim:
     \begin{eqnarray}
     |\psi>\in{\cal H}_0(p_0),\;((\frac{1}{\sqrt{2}} nq_0|
     \psi> = 0,\;\mbox{all}\;n\in{\cal Z} \Longrightarrow
     |\psi> = 0.
     \end{eqnarray}

\noindent
This is so even though again $|\frac{1}{\sqrt{2}} nq_0)) \;{\not\in}
\;{\cal H}_0(p_0)$.  Thus band limited $|\psi>$ are (possibly over)
determined by the overlaps $((\frac{1}{\sqrt{2}} nq_0|\psi>$
of $|\psi>$ with a discrete sequence of (normalized!) coherent 
states, provided $p_0\leq 2\pi/q_0$.

We can now see that in this form the ST permits an extension
to all vectors in ${\cal H}$, using the properties (6.3) of 
the standard coherent states.  We define a sequence of pairwise
orthogonal subspaces ${\cal H}_m(p_0)\subset {\cal H}$
for all $m\in{\cal Z}$ by:
     \begin{eqnarray}
     {\cal H}_m(p_0) &=& \left\{|\psi>\in {\cal H}|
     \varphi(p)=0\;\mbox{for}\;p\; {\not\in}
     \left[\left(m-\frac{1}{2}\right)p_0,
     \left(m+\frac{1}{2}\right)p_0\right]\right\}
     \subset {\cal H} ,\nonumber\\
     {\cal H}&=& {\sum_{\oplus}\atop m\in{\cal Z}}\;{\cal H}_m(p_0) .
     \end{eqnarray}

\noindent
(Now the meaning of the subscript in ${\cal H}_m(p_0)$ is evident).
Thus $(p_0)$ consists of all off-centre band limited wave
functions $\psi(q)$ such that the centre of the momentum
space interval is shifted from zero to $mp_0$, the width 
remaining $p_0$.  On the one hand one sees easily that the
${\cal H}_m(p_0)$ arise from ${\cal H}_0(p_0)$ by action by
integer powers of the momentum space displacement operator
$U(p_0)$:
     \begin{eqnarray}
     {\cal H}_m(p_0)&=& U(p_0)^m  {\cal H}_0 (p_0)\nonumber\\
     &=& U(m p_0) {\cal H}_0 (p_0)\;,\; m\in{\cal Z}
     \end{eqnarray}

\noindent
And on the other hand each ${\cal H}_m(p_0)$ is invariant 
under action by $S_2$ as well as by $S^{-1}_2$.  Moreover when
restricted to any ${\cal H}_m(p_0)$ (or any direct sum
of them over a finite range of $m$ values), both these operators 
remain bounded.  It is also clear that under the action by

$U(m p_0)$ we have the twin results:

     \begin{mathletters}

     \begin{eqnarray}
      U(mp_0)|\frac{1}{\sqrt{2}} nq_0))&=&
      |\frac{1}{\sqrt{2}}(nq_0 + imp_0))) ,\nonumber\\
     ((\frac{1}{\sqrt{2}} n q_0 |U(m p_0)^{-1}&=&
     ((\frac{1}{\sqrt{2}} (nq_0 + i m p_0) |;\\
      |\psi>\in{\cal H}_0(p_0) &\Longleftrightarrow &
      U(m p_0) |\psi>\in {\cal H}_m(p_0) .
      \end{eqnarray}
      \end{mathletters}

\noindent
We can now transfer the statement (6.11) of the ST from 
${\cal H}_0(p_0)$ to each ${\cal H}_m(p_0)$ individually:
     \begin{eqnarray}
     |\psi>\in {\cal H}_m(p_0),\; ((\frac{1}{\sqrt{2}}
     (nq_0 + i m p_0)|\psi > =0,\;\mbox{all}\;
     n\in{\cal Z}\Longrightarrow |\psi > =0.
     \end{eqnarray}

\noindent
In other words such a band limited $|\psi>$ is (possibly over) 
determined by its inner products with the standard
coherent states $|\frac{1}{\sqrt{2}} (nq_0 + i m p_0))$) 
keeping $m$ fixed and taking all $n\in{\cal Z}$.  Once again
we appreciate that this is so even though these coherent
states are not in ${\cal H}_m(p_0)$.

To pass from ${\cal H}_m(p_0)$ to ${\cal H}$ is quite easy.  
We define the projection operators $P_m(p_0)$ onto the 
various orthogonal subspaces ${\cal H}_m(p_0)$ with
standard properties:
     \begin{eqnarray}
      P_m(p_0) &=& \int\limits^{\left(m+\frac{1}{2}\right)p_0}
     _{\left(m-\frac{1}{2}\right) p_0} dp\;|p) (p|\nonumber\\
     &=&U(m p_0)\;P_0(p_0)\;U(m p_0)^{-1} ;\nonumber\\
     P_{m^{\prime}}(p_0)\;P_m(p_0) &=&
     \delta_{m^{\prime}m}\;P_m(p_0) ;\nonumber\\
     P_m(p_0)\;S_2 &=& S_2\;P_m(p_0).
     \end{eqnarray}

\noindent
Then the content of the original ST is fully equivalent to the
following:
     \begin{eqnarray}
     |\psi>\in {\cal H},\;((\frac{1}{\sqrt{2}}
     (n q_0 + i m p_0)| P_m(p_0)|\psi> &=& 0,\;\mbox{all}\;
     m,n\in{\cal Z}\Longrightarrow |\psi> = 0,\nonumber\\
     p_0&\leq & 2\pi/q_0 .
     \end{eqnarray}

It is worthwhile exploring a little bit the real meaning
of implication statements such as eqns.(6.8,11,15,17) in
the following manner.  The subtleties mainly arise from
the use of nonorthonormal systems of vectors as `bases'
in infinite dimensional Hilbert space.  If one has a
complete orthonormal basis $\{|e_n>, \;n=1,2,\ldots\}$ 
for a Hilbert space ${\cal H}$, then any vector 
$|\psi>\in{\cal H}$ has well defined projections 
$<e_n|\psi>$ on to these basis vectors; and the expansion 
of $|\psi>$ in terms of $|e_n>$ with these projections as 
coefficients indeed converges to $|\psi>$ in norm.  The
inclusion of more and more terms in the expansion
improves the accuracy with which $|\psi>$  is approximated,
while in the process the coefficients of already included
terms suffer no change.  Moreover the vanishing of 
$<e_n|\psi>$ for all $n$ implies the vanishing of $|\psi>$.
Lastly we can in principle choose each projection $<e_n|\psi>$ 
independently as we wish, provided that the norm of $|\psi>$
is kept finite.

If we now replace the orthonormal basis $\{|e_n>\}$ by a
nonorthonormal one, $\{|f_n>\}$ say, which may in particular
be overcomplete, the statements that can be made get modified.
In general, the inner products $<f_n|\psi>$ may not be
specifiable independently of one another (over completeness of
$\{|f_n>\}$).  On the other hand the vanishing of all
$<f_n|\psi>$ indeed implies the vanishing of $|\psi>$
(totality of $\{|f_n>\}$).  This means that the closure 
of the set of all finite linear combinations of the
$|f_n>$ is the total space ${\cal H}$.  However, even given 
all these properties, there may be no definite set of 
expansion coefficients with whose help $|\psi>$, in
general, can be expressed as a convergent linear combination 
of the $|f_n>$.  (Over) completeness of $\{|f_n>\}$ will
ensure that any $|\psi>$ can be approximated as closely as
desired via finite linear combinations of the $|f_n>$; but
`in the limit' there may be no `actual expansion' for
$|\psi>$ in terms of $|f_n>$.  Vectors $|\psi>$ in ${\cal H}$
expressible as finite linear combinations of the
$|f_n>$ or as infinite convergent linear combinations with
well-defined expansion coefficients will form a dense
subset in ${\cal H}$.  This situation is well known in
the theory of nonharmonic Fourier series \cite{15}.
It has also been
analysed to a considerable extent in the case of the von
Neumann lattice of standard coherent states, clarifying 
the meaning of expansions of vectors in terms of them or 
of their dual basis vectors\cite{16}.

Keeping all these subtleties in mind, let us agree to use the 
word `basis' in a broad sense for a general possibly over 
complete set of possibly nonorthonormal vectors in 
${\cal H}$.  Then the final result of the original ST of
eqns.(5.5,5.11,6.11) is:
     \begin{mathletters}
     \begin{eqnarray}
     \left\{P_m(p_0)\big|\frac{1}{\sqrt{2}}\left(n q_0 +
     i m p_0\right))),\;n\in{\cal Z},\; m\;\mbox{fixed}\right\}
      &=& \mbox{basis for}\; {\cal H}_m(p_0);\\
     \left\{P_m(p_0)\big|\frac{1}{\sqrt{2}}
     \left(n q_0 + i m p_0\right) )),\;
     n,m\in{\cal Z}\right\} &=&
     \mbox{basis for}\;{\cal H} ;\nonumber\\
     p_0 &\leq & 2\pi/q_0 .
     \end{eqnarray}
     \end{mathletters}

\noindent
It has led to a basis for ${\cal H}$ by setting up bases for each
${\cal H}_m(p_0)$ in turn, and then taking the union over
$m\in{\cal Z}$.

At this point we turn to the well known results concerning
lattices of standard coherent states, which have been mentioned
earlier.  These lattices consist of the vectors 
$|\frac{1}{\sqrt{2}}\left(n q_0 + i m p_0\right) ))$
with $n, m\in{\cal Z}$ and $p_0\leq 2\pi/q_0$.  For
$p_0=2\pi/q_0$ we have the von Neumann lattice, while for
$p_0<2\pi/q_0$ we have a finer lattice.  Then we have
the result\cite{7}
     \begin{eqnarray}
     |\psi>\in {\cal H},\;((\frac{1}{\sqrt{2}}
      \left(nq_0 + i m p_0\right)|\psi> = 0,\;
      \mbox{all}\;m,n\in{\cal Z}\Longrightarrow |\psi>=0 .
      \end{eqnarray}

\noindent
Thus the von Neumann (or any finer) lattice forms a basis for
${\cal H}$.  At $p_0=2\pi/q_0$ (von Neumann Case) we have
over completeness by one vector; while for $p_0<2\pi/q_0$
removal of any finite set of vectors from the lattice does not
destroy over completeness.  Of course for coarser lattices, 
$p_0>2\pi/q_0$, totality is lost.

We can now appreciate how tantalisingly close the statements
based on the ST and on the well known quantum mechanical theory
of coherent state lattices are to one another. The former 
leads to the twin statements (by virtue of symmetry between 
$\hat{q}$ and $\hat{p}$):
     \begin{eqnarray}
     \left\{P_m(p_0)\big|\frac{1}{\sqrt{2}} \left(n q_0+i m p_0\right)
     )),\;m,n\in{\cal Z}\right\}&=&
     \mbox{basis for}\;{\cal H},\nonumber\\
     \left\{\tilde{P}_n(q_0)\big|\frac{1}{\sqrt{2}}
     \left(nq_0 + i m p_0\right))),\;
     m,n\in {\cal Z}\right\}&=&
     \mbox{basis for}\;{\cal H},\;
     q_0p_0\leq 2\pi ,
     \end{eqnarray}

\noindent
where the new projection operators $\tilde{P}_n(q_0)$ are
defined analogously to eqn.(6.16):
     \begin{eqnarray}
     \tilde{P}_n(q_0) = \int\limits^{\left(n+\frac{1}{2}\right)
     q_0}_{\left(n-\frac{1}{2}\right)q_0} dq \;|q><q| .
     \end{eqnarray}

\noindent
The latter leads to the statement
     \begin{eqnarray}
     \left\{\big|\frac{1}{\sqrt{2}}\left(n q_0+i m p_0\right) )),\;
     m,n\in{\cal Z}\right\} =\mbox{basis for}\;{\cal H}.
     \end{eqnarray}

\noindent
These are two distinct properties possessed by the same lattices 
of standard coherent states.  It may not be out of place to 
mention that all the results flowing from the ST are ultimately
based on the properties of the Fourier transformation, while the 
results concerning von Neumann or finer standard coherent state
lattices are generally derived by apealing to the sophisticated 
theory of entire analytic functions, and relations between 
their orders and types and distributions of zeroes.

\section{The ST and lattice systems of H-W Generalized 
Coherent States}

We have seen how to express the ST in the language of standard
coherent states, and how close the results are to earlier
results pertaining to certain phase space lattices of the
latter.  Now, as mentioned in the Introduction and as seen in
Section III in an idealized sense for the Zak basis vectors 
$|q,p>$, the standard coherent states have been extended to 
systems of generalized coherent states (GCS) associated 
with the H-W group, obtained by replacing the Fock ground
state $|0))$ in eqn.(6.1) by a general normalised
fiducial vector $|\psi_0>\in{\cal H}$\cite{2,5}.
 Since the ST in
itself does not refer to any coherent state system at all, 
it is natural to ask if its content could be expressed in 
terms of certain lattices of suitably chosen GCS systems as
well.  We shall find that this can sometimes be done.
This Section will explore the interrelations between
H-W GCS systems, von Neumann type and finer lattices of such
systems, the Zak representation and the ST.  The new terms 
appearing here will be defined as we proceed.  While for 
completeness some old results will be briefly recapitulated
and sometimes sharpened, we will arrive at several  
new insights and results as well.  As we shall
throughout be concerned with the H-W group, continual
reference to this group will be avoided.

Let $|\psi_0>\in{\cal H}$ be a general normalised fiducial
vector, with Schr\"odinger, momentum and Zak wavefunctions
$\psi_0(q), \varphi_0(p), \chi_0(q,p)$ respectively.
(Remember that the last of these depends on the 
parameter $q_0$).  The system of GCS based on $|\psi_0>$,
referred to as $\psi_0$ - GCS hereafter, is defined as the 
family of normalised vectors
     \begin{eqnarray}
      |q^{\prime},p^{\prime};\psi_0> = D(q^{\prime},p^{\prime}) 
       |\psi_0>,\; (q^{\prime},p^{\prime})\in{\cal R}^2.
      \end{eqnarray}

\noindent
It is a well known result that for any choice of $|\psi_0>$,
the $\psi_0$ - GCS family is total, ie. (over) complete in
${\cal H}$\cite{17}.  This is a consequence of the square 
integrable property of the unique UIR of the H-W group.

To obtain the Zak wavefunctions of the $\psi_0$ - GCS, we need
the effect of a general phase space displacement operator
$D(q^{\prime},p^{\prime})$ on a Zak basis vector $|q,p>$.
>From the results in Sections III and IV we find:
     \begin{eqnarray}
     (q,p)\in R(q_0),\;(q^{\prime},p^{\prime})
     \in {\cal R}^2 &:&\nonumber\\
     D(q^{\prime},p^{\prime}) |q,p>&=&
     e^{-i\xi(q,p,-q^{\prime},-p^{\prime})}
     |[q+q^{\prime}], [p+p^{\prime}]>,\nonumber\\
     <q,p| D(q^{\prime},p^{\prime})&=&
     e^{i\xi(q,p,q^{\prime},p^{\prime})}
     < [q-q^{\prime}], [p-p^{\prime}]| ,\nonumber\\
     \xi(q,p,q^{\prime},p^{\prime})&=& qp^{\prime}
      -pq^{\prime}+\frac{1}{2}(qp+q[p-p^{\prime}]
      -p[q-q^{\prime}]\nonumber\\
      &-& [q-q^{\prime}][p-p^{\prime}]).
     \end{eqnarray}

\noindent
Here the fractional parts $[q\pm q^{\prime}], [p\pm p^{\prime}]$
are defined as in eqn.(3.4). We then find that the Zak 
wavefunctions of the vectors in the $\psi_0$- GCS are
given in terms of $\chi_0$ by:
     \begin{eqnarray}
     <q,p|q^{\prime},p^{\prime};\psi_0> =
     e^{i\xi(q,p,q^{\prime},p^{\prime})} \chi_0
    ([q-q^{\prime}], [p-p^{\prime}]) .
    \end{eqnarray}

\noindent
These are thus phase factors times phase space translations
(reduced to or modulo $R(q_0)$) of $\chi_0(q,p)$.

The von Neumann lattice of $\psi_0$ - GCS is the discrete
($q_0$ - dependent) subset of the states (7.1) defined
as follows:
     \begin{eqnarray}
     |n,m;\psi_0> & \equiv & |nq_0, 2\pi m/q_0; \psi_0>\nonumber\\
     &=&(-1)^{mn} U_0^m V_0^n |\psi_0>,\;
     n,m\in{\cal Z}.
     \end{eqnarray}

\noindent
We shall refer to these as the $\psi_0$ - von Neumann GCS
lattice.  Their Zak wavefunctions are naturally simpler than the 
general case in eqn.(7.3):
     \begin{eqnarray}
     <q,p|n,m; \psi_0> = (-1)^{mn}\;
     e^{-i n q_0 p + 2\pi i m q/q_0}
     \chi_0(q,p).
     \end{eqnarray}

\noindent
Naturally no translations of the arguments of $\chi_0$ are 
involved.  Two noteworthy results which have been obtained 
very simply via the Zak description, may be recalled at this 
point\cite{18}:
     \begin{mathletters}
     \begin{eqnarray}
     \{|n,m; \psi_0>\}\;\mbox{total in}\;{\cal H}
     &\Longleftrightarrow & \chi_0(q,p) \neq 0,\;
     (q,p)\in R(q_0) ;\\
     \{|n,m; \psi_0>\}\;\mbox{orthonormal}
     &\Longleftrightarrow &| \chi_0(q,p) | = 1,\;
     (q,p)\in R(q_0) .
     \end{eqnarray}
     \end{mathletters}

\noindent
We see that quite interestingly property (7.6b) implies 
(7.6a):  if the vectors of the $\psi_0$ - von Neumann
GCS lattice are mutually orthogonal, they are also
complete in ${\cal H}$. 

A connection to band limited wave functions may now be
easily seen.  Suppose $|\psi_0>\in{\cal H}_0(p_0)$ for
some $p_0<2\pi/q_0$.  From eqns.(5.3) we know that then
     \begin{eqnarray}
     \chi_0(q,p) = q_0^{-1/2} e^{iqp} \varphi_0(p) ,
     \end{eqnarray}

\noindent
and this certainly does not obey either of
eqns.(7.6).  Thus for such band limited $|\psi_0>$, 
even though the $\psi_0$ - GCS is total, the $\psi_0$ -
von Neumann GCS lattice is neither orthonormal nor total.

Finer lattices of $\psi_0$ - GCS than the von Neumann
lattice are naturally defined in terms of a pair 
$(q_0,p_0)$ obeying $p_0<2\pi/q_0$.  We shall simply
call them $\psi_0$ - finer GCS lattices and define
their elements by:
     \begin{eqnarray}
     |nq_0, m p_0; \psi_0>&=&
     D(nq_0, mp_0) |\psi_0 >\nonumber\\
     &=& e^{imn\;q_0 p_0/2} U(m p_0) V^n_0 |\psi_0>\nonumber\\
     &=& e^{-imn\;q_0 p_0/2} V^n_0 U(m p_0)
     |\psi_0>, \;n,m\in{\cal Z} .
     \end{eqnarray}

\noindent
Since $U(mp_0)$ is now not an integer power of $U_0$,
their Zak wavefunctions are not as simple as in eqn.(7.5).
We shall see that from the ST we can derive some properties
of totality for such finer lattices, analogous to the
results of Section VI.

To proceed in this direction let us recall the way in
which the ST was related to lattices of standard
coherent states in Section VI.  It was by realizing that 
the (ideal) position eigenket $|0>$ and the Fock ground 
state $|0))$ are related by the bounded invertible 
hermitian operator $S_2=e^{-\frac{1}{2}\hat{p}^2}$:
     \begin{eqnarray}
     |0))&=& \sqrt{2} \pi^{1/4} S_2 |0>\nonumber\\
     &=& \frac{1}{\pi^{1/4}}\int\limits^{\infty}
     _{-\infty} dp\;e^{-\frac{1}{2} p^2} |p)
     \end{eqnarray}

\noindent
As is evident, the momentum space wavefunction of
$|0))$ is essentially $e^{-\frac{1}{2}p^2}$
which is (i) square integrable, (ii) bounded and
(iii) nonvanishing for all (finite) $p$.  This gives
us the hint to link up the ST to suitably chosen
lattices of certain $\psi_0$ - GCS systems.

Assume that the fiducial vector $|\psi_0>$ has a momentum 
space wave function $\varphi_0(p)$ which is (of course) square 
integrable, bounded for all $p$, and nonvanishing for 
all (finite) $p$.  It can in general be complex.  Then we
can express $|\psi_0>$ in the following manner.
     \begin{eqnarray}
     |\psi_0> &=& \int\limits^{\infty}_{-\infty} dp
     \varphi_0(p) |p)\nonumber\\
     &=& S\;\int\limits^{\infty}_{-\infty} dp |p)\nonumber\\
     &=&\sqrt{2\pi}\;S |0> ,\nonumber\\
     <\psi_0|&=& \sqrt{2\pi} <0|S^{\dag} ,\nonumber\\
     S&=& \varphi_0(\hat{p}) .
     \end{eqnarray}

\noindent
The similarity to eqn.(7.9) is clear; however $S$ unlike
$S_2$ may not be hermitian.  Now from the properties
assumed for $\varphi_0(p)$ we see that both $S^{-1}$ and
$S^{\dag -1}$, while definable since $\varphi_0(p)$ is
always nonzero, are expected to be unbounded since
$\varphi_0(p)\rightarrow 0$ as $p\rightarrow \pm \infty$.
However, upon restriction to the subspace ${\cal H}_0(p_0)$, 
all the four operators $S, S^{-1}, S^{\dag}, S^{\dag -1}$
are well-defined and leave this subspace invariant.  As
in eqn.(6.9) here we have
     \begin{eqnarray}
     |\psi>\in{\cal H}_0(p_0) \Longleftrightarrow
     S|\psi>, S^{-1} |\psi>, S^{\dag} |\psi>, 
     S^{\dag -1} |\psi> \in {\cal H}_0(p_0) .
     \end{eqnarray}

\noindent
Now we bring in the ST in the form (6.8) and combine it with 
eqns.(7.10,11).  Subject to $p_0\leq 2\pi /q_0$ and since $V_0$ 
commutes with $S$ and $S^{\dag}$, it is equivalent to the
statement 
     \begin{eqnarray}
     |\psi>\in{\cal H}_0(p_0),\; <\psi_0| V_0^n |\psi>=0,\;
     \mbox{all}\; n \in {\cal Z} \Longrightarrow
     |\psi> = 0.
     \end{eqnarray}

\noindent
The vectors $V^n_0|\psi_0>$ are particular elements of the
$\psi_0$ - von Neumann or $\psi_0$ - finer GCS lattice 
defined in eqns. (7.4,8) above:
     \begin{eqnarray}
     V^n_0 |\psi_0> = |n  q_0,\;0;\;\psi_0 >.
     \end{eqnarray}

\noindent
Therefore we can reexpress the ST (7.12) as:
     \begin{eqnarray}
     |\psi>\in {\cal H}_0 (p_0),\;<nq_0, 0; \psi_0|\psi>=0,\;
     \mbox{all}\;n\in{\cal Z} \Longrightarrow
     |\psi> = 0.
     \end{eqnarray}

\noindent
This is a generalisation of (6.11) valid (atleast) when
$\varphi_0(p)$ obeys the stated conditions.  We see
here too, as in Section VI, that even though the vectors
$|n q_0, 0; \psi_0>$ do not belong to ${\cal H}_0(p_0)$,
the overlaps of a band limited $|\psi>\in {\cal H}_0(p_0)$ 
with them are enough to (possibly over) determine 
$|\psi >$.

This result can next be extended to all the subspaces
${\cal H}_m(p_0)$ defined in eqn.(6.12). On the one hand
we have eqn.(6.13) connecting ${\cal H}_0(p_0)$ to
${\cal H}_m(p_0)$.  On the other hand we have from
eqn.(7.8):
     \begin{eqnarray}
     |nq_0, m p_0; \psi_0 > &=&
     e^{i m n q_0 p_0/2} U(m p_0) |n q_0, 0;
     \psi_0 >,\nonumber\\
     <n q_0, m p_0; \psi_0| &=& e^{-i m n q_0 p_0/2}
     <n q_0, 0; \psi_0| U(m p_0)^{-1} .
     \end{eqnarray}

\noindent
Then combining eqns.(6.13,7.15) and the form (7.14) of the ST
we arrive at the statement:
     \begin{eqnarray}
     |\psi>\in {\cal H}_m(p_0),\;<n q_0, m p_0; 
     \psi_0 |\psi > = 0,\;\mbox{all}\;
     n\in {\cal Z} \Longrightarrow |\psi>=0
     \end{eqnarray}

\noindent
This generalises eqn.(6.15) to those fiducial vectors 

$|\psi_0>$ whose momentum space wave functions
$\varphi_0(p)$ are pointwise  nonvanishing and bounded.
Bringing in the projection operators $P_m(p_0)$ onto 
${\cal H}_m(p_0)$ defined in eqn.(6.16), we can give
the extended form of the ST to von Neumann or finer
GCS lattices in ${\cal H}$:
     \begin{eqnarray}
     \varphi_0(p)\;\mbox{normalisable, bounded, pointwise
      nonvanishing}\Longrightarrow\nonumber\\
     \{P_m(p_0) |n q_0, m p_0; \psi_0 >,\;
     n, m \in {\cal Z}\}\;\mbox{total in}\;
     {\cal H}, \;p_0\leq 2\pi/q_0 .
     \end{eqnarray}

We can now summarize our findings.  From the standpoint of the 
ST the ``best statement'' in the direction of totality of
suitable lattices of GCS is given by eqn.(7.17), and here
the presence of the projections $P_m(p_0)$ is unavoidable as
they reflect the band limitedness property basic to the ST
This statement is available for both $p_0< 2\pi/q_0$ 
(finer lattices) and $p_0=2\pi/q_0$ (von Neumann lattices).  
On the other hand, if we ask for the ``best statements'' that 
can be made directly about totality of these lattices,
independent of the ST and avoiding the projections $P_m(p_0)$,
the picture is somewhat complicated.  For $\psi_0$ - von Neumann 
GCS lattices we have the result (7.6a) obtained most effectively
by exploiting the Zak representation.  For $\psi_0$ - finer
GCS lattices there seem to be no comparable general results,
as the Zak representation cannot be easily exploited and we 
have no recourse to the theory of entire functions either.

To all this we must add the remark that boundedness and
pointwise non-vanishing of $\varphi_0(p)$, and pointwise
nonvanishing of $\chi_0(q,p)$, are properties not easily
related to one another.  In the case of the standard 
coherent states, studied in Section VI, both conditions 
happen to be satisfied; and for $p_0< 2\pi/q_0$ the
theory of entire functions comes to our aid.  These remarks
suggest that there are two independent lines of
argument at work here, leading to results of somewhat
divergent characters.

\section{Concluding Remarks}

In this work we have given an account of the interrelations  
between the Poisson Summation Formula and Sampling Theorem on the one hand 
and specific families of coherent state lattices associated with the 
H-W group on the other. In particular, by analysing the content of 
the usual Sampling Theorem from this perspective we are able to arrive at 
certain results on standard coherent state lattices which come pretty 
close to known results on von Neumann and finer standard coherent 
state lattices without recourse to the theory of entire analytic functions. 
We then pursue this line of thought further and show that it enables us to 
make specific statements concerning generalised coherent state lattice 
systems as well. We hope that the unified perspective developed here 
would evidently deepen our understanding of these matters and point the way to 
further interesting developments and generalisations.


\begin{thebibliography}{99}
\bibitem{10} Strictly speaking, this relation does not hold
for arbitrary Lebesgue square integrable wave functions
$\psi(q), \varphi(p)$ since these are defined only upto
sets of measure zero.  One must limit oneself to, say,
the subset of continuous wavefunctions.  These qualifications
will be understood in the sequel.
\end{thebibliography}
\end{document}